\documentclass{aastex62}

\graphicspath{{./}{figures/}}

\received{April -, 2023}
\revised{--,--}
\accepted{--,-}
\submitjournal{ApJ}
\usepackage{lineno}

\shorttitle{Spectral evolution due to hydrodynamic escape}
\shortauthors{Louca et al.}

\begin{document}

\title{Metallicity and Spectral Evolution of WASP-39 b: The Limited Role of Hydrodynamic Escape}

\correspondingauthor{Amy Louca}
\email{louca@strw.leidenuniv.nl}

\author[0000-0002-3191-2200]{Amy J. Louca}
\affil{Leiden Observatory, Leiden University \\
P.O. Box 9513, 2300 RA \\
Leiden,
The Netherlands}

\author{Yamila Miguel}
\affiliation{SRON Netherlands Institute for Space Research, Niels Bohrweg 4, 2333 CA Leiden, The Netherlands}
\affiliation{Leiden Observatory, Leiden University \\
P.O. Box 9513, 2300 RA \\
Leiden,
The Netherlands}

\author{Daria Kubyshkina}
\affiliation{Space Research Institute, Austrian Academy of Sciences, Schmiedlstrasse 6, A-8042 Graz, Austria}

\nocollaboration



\begin{abstract}

The recent observations on WASP-39 b by  JWST have revealed hints of high metallicity within the atmosphere compared to its host star (\citeauthor{Feinstein2022} \citeyear{Feinstein2022}; \citeauthor{Ahrer2022} \citeyear{Ahrer2022}; \citeauthor{Alderson2022} \citeyear{Alderson2022}; \citeauthor{Rustamkulov2022} \citeyear{Rustamkulov2022}; \citeauthor{Tsai2023} \citeyear{Tsai2023}). There are various theories on how these high metallic atmospheres emerge. In this study, we closely investigate the impact of extreme escape in the form of hydrodynamic escape to see its impact on atmospheric metallicity and spectral features such as CH$_4$, CO$_2$, and SO$_2$. We perform a grid simulation, with an adapted version of \texttt{MESA} that includes hydrodynamic escape (\citeauthor{kubyshkina2018} \citeyear{kubyshkina2018}; \citeyear{kubyshkina2020}), to fully evolve planets with similar masses and radii to the currently observed WASP-39 b estimates. 
By making use of (photo-)chemical kinetics and radiative transfer codes, we evaluate the transmission spectra at various time intervals throughout the simulation. Our results indicate that the massive size of WASP-39 b limits the metal enhancement to a maximum of $\sim 1.23$x the initial metallicity. When incorporating metal drag, this enhancement factor is repressed to an even greater degree, resulting in an enrichment of at most $\sim$0.4\%. As a consequence, when assuming an initial solar metallicity, metal-enriched spectral features like SO$_2$ are still missing after $\sim 9$ Gyr into the simulation. This paper, thus,  demonstrates that hydrodynamic escape cannot be the primary process behind the high metallicity observed in the atmosphere of WASP-39 b, suggesting instead that a metal-enhanced atmosphere was established during its formation.

\end{abstract}



\section{Introduction} \label{sec:intro}

With the successful launch of JWST, we are now able to look ever so closely at exoplanet atmospheres. Recent observations of WASP-39 b showed a 26$\sigma$ carbon dioxide detection in the transmission spectrum, a molecule that is thought to appear only in higher metallicity atmospheres. They also found no sign of methane in the atmosphere and a possible hint of sulfur dioxide (\citeauthor{Feinstein2022} \citeyear{Feinstein2022}; \citeauthor{Ahrer2022} \citeyear{Ahrer2022}; \citeauthor{Alderson2022} \citeyear{Alderson2022}; \citeauthor{Rustamkulov2022} \citeyear{Rustamkulov2022}). Both of these findings support the idea of metal enhancement and the latter even photo-chemical effects (\citeauthor{Tsai2023} \citeyear{Tsai2023}) in WASP-39 b's atmosphere. Notably, the host star is thought to have solar-like metallicity (\citeauthor{Polanski2022} \citeyear{Polanski2022}). However, if the planet had the same metallicity as the host star, it would not be expected to have CO$_2$ and SO$_2$ signatures and CH$_4$ depletion without external processes drastically altering the composition. Thus, the question remains as to what processes could explain these features.

There are different scenarios that might explain the increase in the abundance of metals in exoplanet atmospheres relative to the host star. One possible way is through the planet's formation process, where the planet enriches its envelope by accreting solids that dissolve in its atmosphere, leading to an enhanced metallicity compared to that of the star (\citeauthor{Fortney2013} \citeyear{Fortney2013}; \citeauthor{Espinoza2017} \citeyear{Espinoza2017}). Recently, \citet{Khorshif_W39} has shown it is likely that WASP-39b initiated its Type II migration from beyond the CO$_2$ ice line with a possibility of formation within the CO ice line, which could account for its high metallicity through planet formation. A different approach in obtaining a high metallicity atmosphere is through the outgassing of higher mass metals using a diffuse core model, as shown in \citet{Misener}. Finally, another possibility is that the atmospheric escape of lighter particles leads to an enhancement of the metals (e.g., \citeauthor{Chen_2016} \citeyear{Chen_2016}). In particular, extreme particle escape in the form of hydrodynamics is thought to have a great impact on the fractionation between lighter and heavier particles within an atmosphere.  Modeling the evolution of (sub-)Neptune mass planets, \citet{Malsky2020} found significant metallicity enhancement with factors ranging from 2x to 11.5x initial metallicity due to extreme photo-evaporation. The loss of heavy particles, however, has been neglected in their evolution models.

Motivated by these ideas, we investigate in this letter whether the impact of hydrodynamic escape on atmospheric composition could account for the high atmospheric metallicty and spectral features observed in the atmosphere of WASP-39 b, to try to disentangle between these potential explanations. Our study begins with a planet having an atmospheric metallicity similar to its host star, and we incorporate metal drag into our atmospheric escape calculations to trace the change in its composition due to the escape of lighter and heavier elements. We simulate the atmospheric, interior structure and evolution of the planet by making use of the planetary evolution code \textit{Modules for Experiments in Stellar Astrophysics}, \texttt{MESA} (\citeauthor{mesa1} \citeyear{mesa1}; \citeyear{mesa2}; \citeyear{mesa3}; \citeyear{mesa4}; \citeyear{mesa5}; \citeauthor{mesa6} \citeyear{mesa6}), where hydrodynamics is included (\citeauthor{kubyshkina2018} \citeyear{kubyshkina2018}; \citeyear{kubyshkina2020}), also considering the effect of metal drag. We use (photo)chemical kinetics and radiative transfer calculations to determine the compositional and temperature structure of the atmosphere, and how it changes over time, with the ultimate goal of understanding the evolution of spectral features and explain the observations. 

\section{Methodology} \label{sec:methods}

The calculations are done in three main steps. First, the interior and atmosphere of WASP-39 b is evolved in time using \texttt{MESA} (\citeauthor{mesa1} \citeyear{mesa1}; \citeyear{mesa2}; \citeyear{mesa3}; \citeyear{mesa4}; \citeyear{mesa5}; \citeauthor{mesa6} \citeyear{mesa6}), where  hydrodynamic escape is included in the form of a hydrobased approximation (\citeauthor{kubyshkina2018} \citeyear{kubyshkina2018}; \citeyear{kubyshkina2020}), and metal drag of heavy particles in the posterior calculations. 
The output evolutionary tracks of the mass, radius, and metallicity at different selected ages, are subsequently used in an open-source radiative transfer python code, \texttt{HELIOS} (\citeauthor{Malik2017} \citeyear{Malik2017}; \citeyear{Malik2019}), from which the temperature-pressure profile of WASP-39 b's atmosphere is forwarded to the open-source chemical kinetics code, \texttt{VULCAN} (\citeauthor{Tsai17} \citeyear{Tsai17}; \citeyear{Tsai21}) to calculate its chemistry and composition of the atmosphere. These atmospheric simulations will tell us more about how the composition and thermal structures evolve over time. Finally, we make use of another open-source radiative transfer code, \textit{petitRADTrans}, hereafter \texttt{pRT} (\citeauthor{prt2019} \citeyear{prt2019}; \citeyear{prt2020}; \citeauthor{prt2022} \citeyear{prt2022}), to see how these compositional and thermal changes affect the synthetic spectra over time. 

\subsection{Evolution in MESA and atmospheric escape approach}

For the evolution of the planets, we make use of an adapted version of \texttt{MESA} that includes hydrodynamic escape in the form of an analytical approximation (\citeauthor{kubyshkina2018} \citeyear{kubyshkina2018}; \citeyear{kubyshkina2020})\footnote{The data and models are available at:\dataset[zenodo.4022393]{https://doi.org/10.5281/zenodo.4022393}}. We explore two extreme cases to establish both lower and upper bounds on the potential metallicity enhancement: one where the drag of heavy elements is considered and another where metal drag is not taken into account.

The drag of heavy metals due to the extreme escape of lighter particles should be taken into account when calculating the metallicity enhancement (see e.g. \citeauthor{Fortney2013} \citeyear{Fortney2013}). For that, we make use of the \textit{crossover mass} criterion (chapter 5 of \citet{Catling2017}),

\begin{equation}
    m_{\mathrm{c}} = m_1 + \frac{kT F_1}{bgX_1}
\label{eq:mcross}
\end{equation}
where $m_1$ is the mass of the major constituent in the gas (i.e. hydrogen), $k$ is the Boltzmann constant, $T$ is the temperature of the atmosphere, $F_1$ is the vertical flux of the major constituent in the gas that escapes the atmosphere, $b$ is the binary diffusion parameter, $g$ is the gravity acceleration of the planet, and $X_1$ is the mixing ratio of the bulk gas in the atmosphere. If the molecular mass of the minor species is lower than the crossover mass it will be dragged along with the major constituent outside the Roche-lobe radius and the ratio of metals to hydrogen will remain the same. Naturally, this is dependent on both the escaping flux and the mass of the planet. The more extreme the escape and the lower the planetary mass, the more metal drag there is.
Within this study, we look at both the metal drag- and non-metal drag case to get lower- and upper estimate of the metal-enhancement within the atmosphere of WASP-39 b.

We assume that metals like N, C, O, and S are in atomic form in the upper atmosphere. 
As the atomic mass of each species differs, the evolution path of the individual metals should differ as well. In this study, however, we simplify this matter by treating the atomic mass of oxygen as representative of the metals in the atmosphere. We make the assumption that all metals evolve similarly to oxygen. The binary diffusion parameter is taken to be $b = 4.6\cdot10^{19}$ cm$^{-1}$ s$^{-1}$ for a planet with an equilibrium temperature of $T_{\mathrm{eq}} = 1100$ K (following \citeauthor{Marrero2009} \citeyear{Marrero2009}). Since the flux and gravity components evolve over time, the crossover mass changes throughout the evolution as well. Hydrodynamic escape is most prominent in the early stages and then simmers down as the planet evolves, which causes heavy particles to escape mostly at these early stages and metal enhancement is expected to kick in only at the later stages. Metal drag, therefore, suppresses the final metallicity enhancement. For the phase where metal drag is prominent, we assume that the metal to hydrogen fraction remains constant. When the molecular mass of the minor species is equal to- or higher than the crossover mass, the metals are not expected to be dragged along anymore and we assume that all escaping mass is pure hydrogen and helium. As a final assumption, we also keep the hydrogen-to-helium ratio constant throughout the simulation. A more mathematical description of the posterior metal enhancement calculations can be found in appendix \ref{app:metallicity}.

To overcome the large error margin around the currently observed mass estimate of $M = 89\pm 9.5 M_{\oplus}$ (\citeauthor{Faedi2011} \citeyear{Faedi2011}) and the lack of constraint in the envelope fraction, we perform a grid-simulation for evolving WASP-39 b. Within the grid simulation, initial masses range between 89.5 $M_{\oplus}$ - 95.5 $M_{\oplus}$ and initial envelope fractions range between 0.3 - 0.65. For all simulations, we use a simplistic core-atmosphere model where the core is assumed to be rocky (silicates and heavy metals) surrounded by a hydrogen dominated envelope. The envelope is assumed to be well-mixed and homogeneous throughout, without any compositional gradient in the X, Y, and Z fractions and, therefore, with the same composition as the atmosphere. While the experience on solar system giants tells us that this might not hold true for giant exoplanets (e.g. \citeauthor{Bloot2023} \citeyear{Bloot2023}; \citeauthor{Miguel2022} \citeyear{Miguel2022}; \citeauthor{Mankovich2021} \citeyear{Mankovich2021}), a more detailed calculation including compositional gradients is out of the scope of this letter and will be studied in future publications. As initial conditions, we assume the atmosphere to have stellar abundance, which is solar-like for the case of WASP-39 (\citeauthor{Mancini2018} \citeyear{Mancini2018}).
Finally, we note that in this letter we include evolution processes that might inflate the planetary radius (e.g. \citeauthor{MolLous2020} \citeyear{MolLous2020}; \citeauthor{Komacek2017} \citeyear{Komacek2017}). For the lower envelope mass fractions (i.e. $f < 0.6$), we include an internal luminosity of $L_{\mathrm{int}} = 10^{27}$ erg s$^{-1}$, while for the higher envelope fractions ($f \geq 0.6$) the internal luminosity is set to $L_{\mathrm{int}} = 5\cdot10^{26}$ erg s$^{-1}$ to ensure that atmospheric runaway does not occur within the lifetime of the planet. 

\subsection{Compositional and thermal evolution of the atmosphere}

The compositional and thermal structures in the atmosphere are evolved using the open-source python codes \texttt{VULCAN} (\citeauthor{Tsai17} \citeyear{Tsai17}; \citeyear{Tsai21}) and \texttt{HELIOS} (\citeauthor{Malik2017} \citeyear{Malik2017}; \citeyear{Malik2019}), respectively. \texttt{HELIOS} is a one-dimensional, open-source radiative transfer code to calculate temperature-pressure profiles for exoplanet atmospheres. The output temperature-pressure profile is forwarded into \texttt{VULCAN}, which is a one-dimensional chemical kinetics code that includes photo-chemistry.
The initial elemental abundances are assumed to be stellar-like for all elements but O. From recent JWST observations, the C/O is known to be sub-solar (\citeauthor{Alderson2022} \citeyear{Alderson2022}; \citeauthor{Feinstein2022} \citeyear{Feinstein2022}; \citeauthor{Ahrer2022} \citeyear{Ahrer2022}; \citeauthor{Rustamkulov2022} \citeyear{Rustamkulov2022}), and we therefore assume C/O = 0.25 by increasing the oxygen abundance.
As the planet evolves, the metallicity, radius, and mass parameters are updated accordingly to the evolutionary tracks of the modeled planets in \texttt{MESA}. The stellar parameters (i.e. effective temperature and radius) are updated using the MIST database (\citeauthor{Dotter2016} \citeyear{Dotter2016}; \citeauthor{Choi2016} \citeyear{Choi2016}; \citeauthor{mesa1} \citeyear{mesa1}; \citeyear{mesa2}; \citeyear{mesa3}), to simultaneously reflect the stellar evolution. We make use of the PHOENIX models (\citeauthor{phx95} \citeyear{phx95}) to update the stellar spectra. For the chemical kinetics code \texttt{VULCAN}, we make use of the default S-N-C-H-O (photo-)chemical network that includes 1285 forward-, backward-, and photochemical reactions. Vertical mixing is included in the form of molecular- and eddy diffusion. We make use of an eddy diffusion constant of $K_{zz} = 10^{10} \mathrm{\ cm^{2} \ s^{-1}}$, which has been previously adopted by other studies (e.g., \citeauthor{Moses2012} \citeyear{Moses2012}; \citeauthor{Parmentier2013} \citeyear{Parmentier2013}; \citeauthor{Miguel2014} \citeyear{Miguel2014}). For radiative transfer, we include the opacities of the molecules H$_2$O, CH$_4$, CO, H$_2$S, CO$_2$, SO$_2$, PH$_3$, H$_2$ and the atoms H, He, Na, and K. The line lists are taken from the open-source DACE database\footnote{\href{https://dace.unige.ch/opacityDatabase/?\#}{dace.unige.ch}}
(\citeauthor{Grimm2015} \citeyear{Grimm2015}; \citeauthor{Grimm2021} \citeyear{Grimm2021}) and are listed in table \ref{tab:linelists}. The atomic and molecular abundances are initially calculated using the chemical equilibrium code \texttt{FastChem} (\citeauthor{Stock2018} \citeyear{Stock2018}). We make use of a resolution of $\lambda / \triangle \lambda = 1000$ within the wavelength range $\lambda_{\mathrm{min}} = 0.06$ $\mu$m and $\lambda_{\mathrm{max}} = 200$ $\mu$m. We also account for global heat-redistribution on WASP-39 b and a constant internal temperature of $T_{\mathrm{int}} = 350$ K (\citeauthor{Tsai2023} \citeyear{Tsai2023}).

\subsection{Evolution of transmission spectra}

The transmission spectra are computed using the open-source radiative-transfer python code \texttt{pRT} (\citeauthor{prt2019} \citeyear{prt2019}; \citeyear{prt2020}; \citeauthor{prt2022} \citeyear{prt2022} ). We include the opacities of the molecules H$_2$O, CH$_4$, SO$_2$, H$_2$S, CO$_2$, CO. All line lists used for each species are indicated in table \ref{tab:linelists}. We take the default line-list from pRT for all molecules, excluding SO$_2$. For SO$_2$ we make use of the recommended line-list from \href{https://www.exomol.com/}{ExoMol} (\citeauthor{TENNYSON201673} \citeyear{TENNYSON201673}). As scattering species we include H$_2$-H$_2$ and H$_2$-He. The resolution is set to $\lambda/\triangle\lambda \leq 1000$.  
At each time step, we update the abundances, temperature profiles, radius and mass of the planet according to the output of \texttt{VULCAN}, \texttt{HELIOS}, and \texttt{MESA}.  

\section{Results} \label{sec:results}

\subsection{Metal evolution}

\begin{figure}
    \centering
    \includegraphics[scale=0.9]{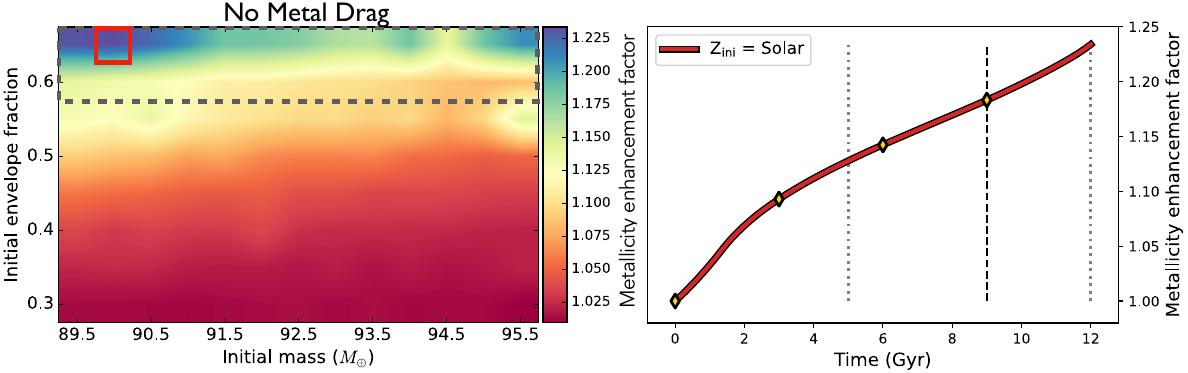}
    \caption{The metallicity enhancement after $\sim 12$ Gyr of the WASP-39 b grid simulations when solar initial metallicity is assumed (left) and the evolution of the metal-enhancement factor over time (right) for one specific case when solar initial metallicity is assumed and no metal drag is included (red line). The evolution of the metallicity enhancement factor is plotted for the most extreme case, indicated by the red box in the left plot. For the grid plot (left), the color represents the metal-enhancement factor with respect to the initial metallicity, and the horizontal and vertical axes represent the initial mass and initial envelope fraction respectively. Note that the grid plot is smoothed using the default anti-aliasing interpolation function. The dashed grey rectangle represents the region within the grid plot where a lower internal luminosity of $L_{\mathrm{int}} = 5\cdot10^{26}$ erg s$^{-2}$ is assumed. For all simulations outside this region, we assume a higher internal luminosity of $L_{\mathrm{int}} = 10^{27}$ erg s$^{-2}$. The maximum metal enhancement for the solar initial metallicity case is $\sim 1.23$x initial metallicity. The vertical dashed black line in the evolution plot is the current age estimation of the WASP-39 system (\citeauthor{Bonomo2017} \citeyear{Bonomo2017}), with error bars as indicated by the vertical dotted grey lines. The yellow diamonds in the evolution plot represent the time steps at which the transmission spectra are evaluated.}
    \label{fig:MESA}
\end{figure}

The evolved metal enhancements for WASP-39 b without considering metal drag are shown in figure \ref{fig:MESA}.
We evolve the planet for 12 Gyr to account for the maximum age allowed by the error bars in the age determination (\citeauthor{Faedi2011} \citeyear{Faedi2011}). All grid points match WASP-39 b's mass, and with the inclusion of additional internal energy flux, they also fit its radius within the respective error bars for mass, radius, and age of the system. The ratio between the final and initial metallicity, or metal-enhancement factor, for all grid points at the end of the evolution can be seen in the left plot of figure \ref{fig:MESA}. The color gradient from low to high initial envelope fractions within the grid plot is prominent. Metal enhancement is allegedly strongest for  planets with a high initial envelope fraction. This is due to the larger radius that is obtained with bigger envelopes as the gravity component becomes smaller when having same initial masses. As shown in \citet{kubyshkina2018}, a larger radius results in a higher mass loss, $\triangle M$, and, thus, a bigger increase in metallicity (see eq. \ref{eq:metal_enhance}). Additionally, there is a modest color gradient from lower to higher initial masses, though less noticeable\footnote{since the gravity component is more sensitive to changes in radius, i.e. $g\propto \frac{M}{R^2}$} due to the small relative mass range used within the grid simulation. For planets with higher mass, the metal enhancement is less pronounced. This is also primarily attributed to the gravity component, which increases with mass.

The evolution of metallicity within the atmosphere of the most extreme case (indicated by the red box in the grid-plot of figure \ref{fig:MESA}) is shown in the right plot of figure \ref{fig:MESA}. 
When assuming solar initial metallicity, the upper limit of metal enhancement within these grid points is $\sim 23$\%.

We see that not even in the most extreme case without metal drag the enhancement in metallicity is big enough to explain the observed metallicity of the planet when starting with stellar metallicity. The WASP-39 b observed enrichment is, thus, not solely due to atmospheric escape but to a combination of formation and evolution through atmospheric escape (see section \ref{sec:discussion}).

\subsection{Spectral evolution}

\begin{figure}[t]
    \centering
    \includegraphics[scale=0.9]{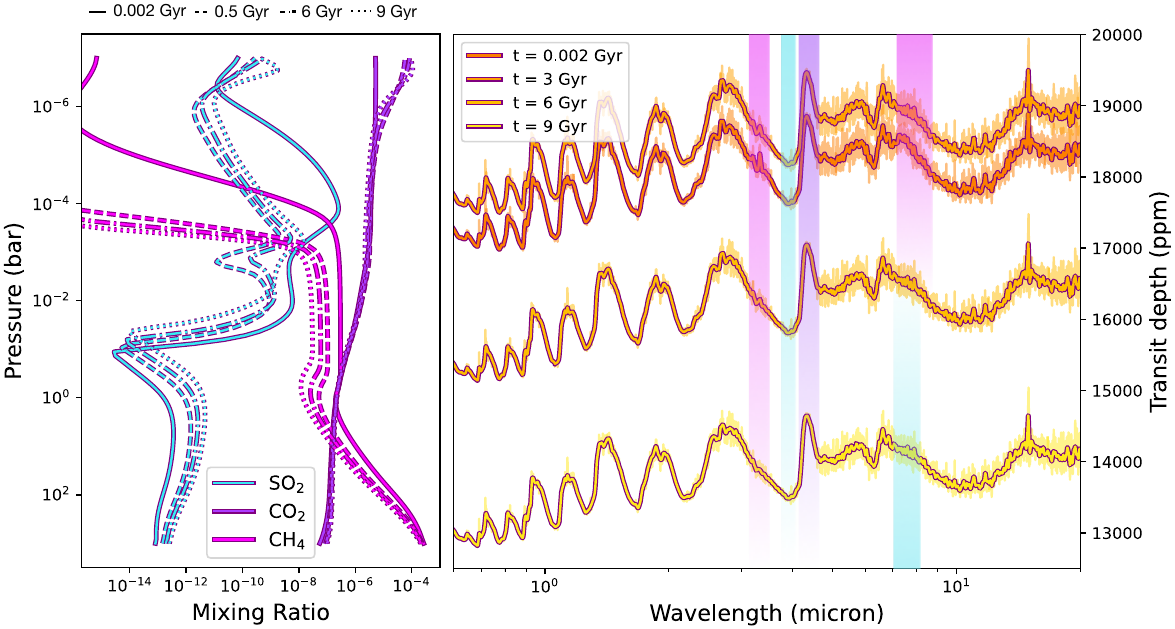}
    \caption{The evolving mixing ratios (left) and transmission spectra (right) of WASP-39 b for the most extreme metal-enhanced scenario, at different time steps in the simulation. The time step is shown within each transmission evolution plot (right). Highlighted in each plot are CH$_4$ features at 3.3 microns and 7-8 microns, CO$_2$ features around 4-5 microns, and missing SO$_2$ features around 4 microns and 7-8 microns, colored similarly as in the mixing ratio plot. The solid lines in the mixing ratio plots represent the abundances at 0.002 Gyr, the striped lines at 0.5 Gyr, the dotted-striped lines at 6.0 Gyr, and the dotted lines at 9.0 Gyr, as labelled above the left figure. The time stamps for the transmission spectra plot are labeled within the figure.}
    \label{fig:trans_evo}
\end{figure}

Figure \ref{fig:trans_evo} illustrates the chemical (left) and spectral (right) evolution of WASP-39 b due to hydrodynamic escape in the extreme case when metal drag is not taken into account and when starting the simulations assuming stellar metallicity. At first glance, the changes in spectra seem insignificant. At each stage, the \textbf{SO$_2$} feature around 4 microns as well as 7-8 microns, is absent, which is expected since the metallicity reached here is not big enough to show that feature (\citeauthor{Polman2023} \citeyear{Polman2023}). We see from the left plot in figure \ref{fig:trans_evo} that the maximum mixing ratio of SO$_2$ is around $\sim 10^{-7}$ between $10^{-4}-10^{-6}$ bar. This value is about 1-2 orders of magnitude too low to be observed for SO$_2$, as argued in \citet{Polman2023} and \citet{Tsai2023}. The SO$_2$ feature at 7-8 microns is also absent. Instead, we see an CH$_4$ feature at 7-8 microns. At the same time, another moderate CH$_4$ feature is found around 3.3 microns. These features are in agreement with the relatively high CH$_4$ abundance in the upper atmosphere (at \(P \approx 10^{-4}\) bar) observed at young ages (0.002 Gy). The high CH$_4$ abundance at \(t = 0.002\) Gyr is caused by the young host star, which still maintains a relatively low temperature at this young age. This, in turn, reduces the equilibrium temperature of WASP-39 b, facilitating the persistence of CH$_4$ at these pressure levels. This opens a potential window to observe CH$_4$ on young Saturn-sized planets similar to WASP-39 b (by using e.g., the MIRI instrument of JWST for the 7-8 microns feature). Both enhanced CH$_4$ features quickly dissipate between 0.002 and 3 Gyr. Note that at these stages, the metallicity increases by only a factor of 1.1x, and the changes in the spectra can be attributed to stellar evolution. As a result of the slight increase in metallicity and increase of stellar temperature, the minor CH$_4$ feature gradually dissipates after 6 Gyr. 
Contrarily, the \textbf{CO$_2$} features at 4-5 and 10-20 microns is present throughout the entire evolution. These spectral features are consistent over time and show no observable change.

From the transmission spectra, it can be seen that the radius of the planet decreases due to atmospheric escape, with the exception of the period between 0.002 - 3 Gyr. Here we see an increase in transit depth which is caused by the evolution of the stellar radius that decreases over time. After 3 Gyr, the transit depth becomes less significant over time, going from $\sim 19000$ ppm to $\sim 14000$ ppm. This can be fully attributed to atmospheric escape as the internal luminosity restrains the planet from contracting and the stellar radius shows little to no change at this age.

\subsection{The effect of metal-drag}

\begin{figure}
    \centering
    \includegraphics[scale=0.85]{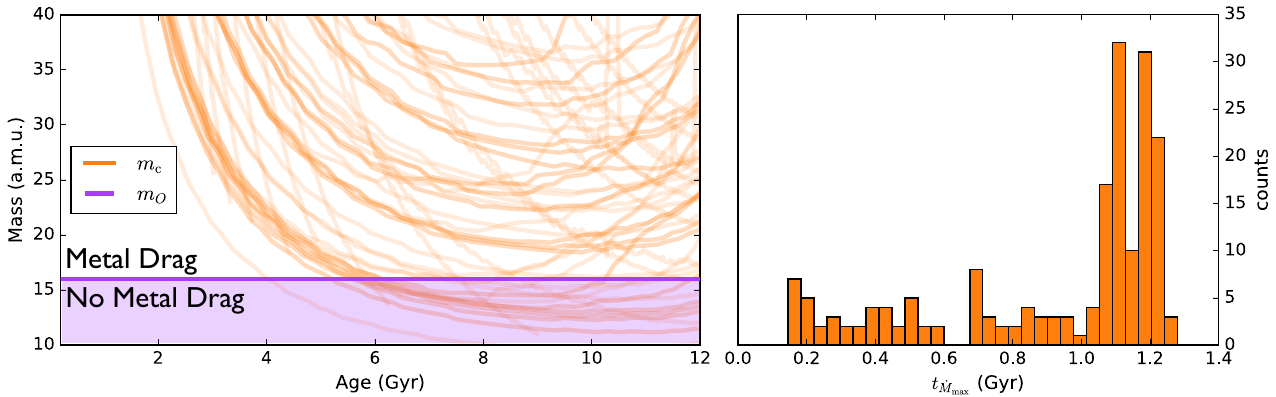}
    \caption{The evolution of the crossover mass for all simulations (left) and the distribution of the age at which the gradient in mass, $t_{\dot{M}_{\mathrm{max}}}$, is the largest (right). In the evolution plots, the orange lines represent the crossover masses of the simulations while the horizontal purple line represents the atomic mass of oxygen in amu. This horizontal line embarks the threshold for which metal drag should occur.}
    \label{fig:mcross}
\end{figure}

In this study, metal drag is included when the crossover mass lies above the mass of metals, as shown in figure \ref{fig:mcross} (left). The figure shows that the crossover mass of most planets always exceeds the chosen metal mass. This tells us that these metals will be dragged along with extreme hydrogen and helium escape throughout the evolution of these planets. However, we still see a fair amount of planets that fall below this threshold line after a few Gyr, indicating that metal enhancement should take place on these planets. The maximum metallicity enhancement factor found for those cases is $\sim$0.4\%. The reason for such a small metallicity enhancement is the high age at which the crossover masses have settled down. The right plot of figure \ref{fig:mcross} shows the distribution of the ages at which we have maximum mass outflow, $\dot{M}_{\mathrm{max}}$, which is at maximum $\sim$ 1.2 Gyr for most planets in the grid simulation. After this period, the atmospheric outflow relaxes as the planet stabilizes (until the re-inflation period, as described in \citet{Thorngren2021}). The age distribution peaks at around $\sim$1.2 Gyr, while in the left plot of figure \ref{fig:mcross}, we see that the age at which the crossover masses have relaxed enough is $\geq$4 Gyr. As the metal-enhancement is highly dependent on the atmospheric mass loss (see equation \ref{eq:metal_enhance}), we do not expect to see substantial metal-enhancement after $\sim$ 1.1 Gyr into the simulation. Taking this into account, metal drag will cause little to no metal enhancement for all planets considered.

\newpage
\section{Discussion}
\label{sec:discussion}
In the previous section, we showed that the high metallic spectral features found in the atmosphere of WASP-39 b from recent JWST observations are not the result of the extreme escape of lighter particles when starting the planetary evolution from a planet with stellar metallicity. SO$_2$ features, specifically, are challenging to obtain from the low metal enhancement obtained due to hydrodynamic escape in this case. Metal drag and the large size of WASP-39 b are restraining factors to the metal enhancement in the evolution, which, consequently, also confines any sulfur dioxide features.

A different approach to obtain a high-metallicity atmosphere is to think that the current metallicity of the WASP-39 b's atmosphere is the result of formation processes. Starting with a metal enriched atmosphere alters planetary evolution. The most obvious change in evolution is the decreasing radius due to the increasing mean molecular mass. The gravitational pull will, consequently, be bigger as $g \propto R^{-2}$, which makes it more challenging for atmospheric particles to reach escape velocity. The final metal enhancement will then be slowed down and we expect the enhancement factor to be less than 23\%, as previously found in this study.

Another possibility to enhance the metallicity of the planet is through the dilution of the core and subsequent enrichment of the atmosphere (\citeauthor{Misener} \citeyear{Misener}). Furthermore, in this paper we assumed that the planet consists of a core surrounded by a homogeneous envelope, but an internal structure with compositional gradients is favoured by formation models on giant planets (\citeauthor{Helled2022} \citeyear{Helled2022}; \citeauthor{Valletta2019} \citeyear{Valletta2019}; \citeauthor{Lozovsky2018} \citeyear{Lozovsky2018}; \citeauthor{Brouwers2018} \citeyear{Brouwers2018}; \citeauthor{Venturini2016} \citeyear{Venturini2016}; \citeauthor{Hori2011} \citeyear{Hori2011}) and by recent models and observations on the giant planets in our solar system (\citeauthor{Miguel2022} \citeyear{Miguel2022}; \citeauthor{Mankovich2021} \citeyear{Mankovich2021}; \citeauthor{Wahl2017} \citeyear{Wahl2017}). In that case, the evaporation of H and He in the atmosphere would expose more enriched layers, which is another potential venue for enrichment due to evolution in Saturn-mass planets like WASP-39 b.

Finally, we also showed an enhanced CH$_4$ feature in the early stages of the planet, explained by the relatively low equilibrium temperatures in this phase. A caveat in this study is that we have made use of stellar spectra from the PHOENIX models that does not take the increased UV-flux of the young host star into account. Previous studies have shown, however, that increased UV-flux is expected for younger stars (e.g., \citeauthor{Claire2012} \citeyear{Claire2012}). It is likely that an enhanced UV-flux will dissociate CH$_4$ into smaller particles and, therefore, reduce the CH$_4$ feature in this adolescent phase.

\section{Conclusion}
\label{sec:conclusion}

In this study, we investigated the impact of hydrodynamic escape on the evolution of metallicity and spectral features of WASP-39 b to determine if this process can explain current JWST observations. We made use of a planetary evolution code for simulating a grid population of planets similar to WASP-39 b to get the time-dependent radius, mass, and metallicity. For several time steps, these parameters were then forwarded to a radiative transfer code and a (photo-)chemical kinetics code to obtain the time-dependent transmission spectra based on the evolving temperature-pressure profile and chemical composition of the atmosphere.

We found that, the most extreme case of metal enrichment due to particle escape shows a metal enhancement of $\sim 23.3$\% after $\sim 12$ Gyr. This enhancement factor does not account for metal drag and, therefore, puts an upper limit to metal enrichment for WASP-39 b due to hydrodynamic escape. For this specific case, the evolving transmission spectra did not show any significant difference over time. After $\sim 9$ Gyr into the simulation, no SO$_2$ features were present in the spectrum and the CO$_2$ feature remained approximately the same. The biggest spectral changes were found to be CH$_4$ features that appeared after $\sim 0.002$ Gyr into the simulation at 3.3 and 7-8 microns and disappeared again after $\sim 6$ Gyr into the simulation - possibly opening a time-window of the presence of CH$_4$ in young gaseous exoplanet atmospheres like WASP-39 b.  The CH$_4$ features were caused by the change in stellar temperature due to stellar evolution. This can be tested in future observational endeavors that focus on comparing Saturn-mass adolescent planets with more mature ones.

We also showed that metal drag suppresses the metal enhancement significantly, leaving a maximum enrichment of $\sim$0.4\%. The metal enhancement was mainly restrained due to the fact that metal drag lasted throughout the most extreme atmospheric escape period (i.e. $\leq 1.2$ Gyr).

All in all, using hydrodynamic escape models we cannot explain the observed features of WASP-39 b assuming that the planet was formed with a stellar metallicity, since the evolution of the planet would not account for the observed features in this case, putting strong constraints on planet formation processes.

\appendix

\section{Metal enhancement}
\label{app:metallicity}

The \textit{metal enhancement factor} (i.e. $Z/Z_{\mathrm{ini}}$, where $Z$ is the current metallicity and $Z_{\mathrm{ini}}$ is the initial metallicity) is calculated in a posterior manner using iterative equations. After the evolution of the planet within \texttt{MESA}, the time-dependent mass is used to calculate the metal enhancement factor. This is done by assuming that 1) the mass loss consists only of hydrogen- and helium particles, and 2) the hydrogen-to-helium fraction remains constant throughout the simulation. The current metallicity is evaluated at each time step by calculating the time-dependent hydrogen- and helium fractions, $X$ and $Y$ respectively

\begin{equation}
\label{total_content}
    Z_{i+1} = 1 - X_{i+1} - Y_{i+1}
\end{equation}

here $X$, $Y$, and $Z$ are the \textit{mass fractions} of the hydrogen- helium- and heavy metals content, which, in mathematical terms, can be described as

\begin{equation}
\label{mass_fraction}
    A_{i+1} = \frac{M_{A,i+1}}{M_{\mathrm{tot},i+1}}
\end{equation}
where $A$ is the atom-type (i.e. $X$, $Y$, or $Z$) at time step $i+1$. The hydrogen- and helium fractions are iteratively calculated by monitoring the mass of the planet at each time step and, from that, calculating the mass loss

\begin{equation}
\triangle M_{i,i+1} = M_{\mathrm{tot},i} - M_{\mathrm{tot},i+1}
\end{equation}
where $M_i$ and $M_{i+1}$ are the mass of the planet at time steps $i$ and $i+1$ respectively. This mass can be fully attributed to hydrogen- and helium mass loss. Since $X + Y < 1$, we need an extra multiplication factor to distribute this mass loss component over only hydrogen and helium escape, i.e.

\begin{equation}
    B_i(X_i\cdot \triangle M_{i,i+1} + Y_i\cdot \triangle M_{i,i+1}) = \triangle M_{i,i+1}
\end{equation}
which gives for $B$
\begin{equation}
    B_i = \left(\frac{1}{X + Y}\right)_i
\end{equation}
The hydrogen- and helium masses at time step $i+1$ can now be calculated by subtracting the hydrogen/helium mass loss at time step $i+1$ from the current hydrogen/helium mass,

\begin{equation}
\label{hydrogen_content}
    M_{\mathrm{A},i+1} = A_i \cdot M_{\mathrm{tot},i} - \left(A\cdot B\right)_i\cdot \triangle M_{i,i+1}
\end{equation}

By combining equation \ref{mass_fraction} with equation \ref{hydrogen_content}, we finally get an expression for the time-dependent hydrogen- and helium mass fractions
\begin{equation}
    A_{i+1} = \frac{ A_i \cdot M_{\mathrm{tot},i} - \left(A\cdot B\right)_i\cdot \triangle M_{i,i+1}}{M_{\mathrm{tot},i+1}}
    \label{eq:metal_enhance}
\end{equation}
where $A$ now is either the hydrogen- or helium mass fraction (i.e. X or Y respectively). From this, we can obtain the time-dependent metallicity factor by substituting this expression back into equation \ref{total_content}.

Metal drag is included in the form of a step function, meaning that we take it into account when the crossover mass is larger than the mass of the secondary, heavier species. When this is satisfied, we simply use $A_{i+1} = A_{i}$, and when the crossover mass is smaller than the secondary species we use the expression of equation \ref{eq:metal_enhance}.

\newpage
\section{Line lists}
\label{app:linelists}

\begin{table}[ht]
    \centering
    \begin{tabular}{c|c|c|c|c}
        \textbf{Molecule} & \textbf{Temperature range (K)} & \textbf{Pressure range (bar)} & \textbf{Line lists \texttt{HELIOS}} & \textbf{Line lists \texttt{pRT}}  \\
        \hline
        \hline
        H$_2$O & 50 - 2900 & $10^{-8}$ - $10^3$ & POKAZATEL$^1$ & POKAZATEL$^1$ \\
        CH$_4$ & 50 - 2900 & $10^{-8}$ - $10^3$ & YT34to10$^2$ & YT34to10$^2$ \\
        SO$_2$ & 50 - 2900 & $10^{-8}$ - $10^3$ & ExoAmes (v2)$^3$ & ExoAmes (v2)$^3$\\
        CO & 50 - 2900 & $10^{-8}$ - $10^3$ & Li2015$^4$ & HITEMP$^{11}$ \\
        H$_2$S & 50 - 2900 & $10^{-8}$ - $10^3$ & AYT2$^5$ & AYT2$^5$\\
        CO$_2$ & 50 - 2900 & $10^{-8}$ - $10^3$ & CDSD-4000$^6$ & UCL-4000$^{12}$\\
        PH$_3$ & 50 - 2900 & $10^{-8}$ - $10^3$ & SAlTY$^7$ & N/A \\
        H$_2$ & 50 - 2900 & $10^{-8}$ - $10^3$ & RACPPK$^8$ & N/A \\
        \hline
        \hline
        \textbf{Atom} & & & & \\
        \hline
        \hline
        H & 2500 - 6100 & $10^{-8}$ & VALD$^9$ & N/A\\
        He & 2500 - 6100 & $10^{-8}$ & Kurucz$^{10}$ & N/A\\
        Na & 2500 - 6100 & $10^{-8}$ & Kurucz$^{10}$ & N/A\\
        K & 2500 - 6100 & $10^{-8}$ & Kurucz$^{10}$ & N/A\\
    \end{tabular}
    \caption{The line lists used for each opacity species included in radiative transfer. [1] \citet{Polyansky2018}; [2] \citet{Yurchenko2014}; \citet{Yurchenko2017} [3] \citet{Underwood2016}; [4] \citet{Li_2015}; \citet{Somogyi2021}; [5] \citet{Azzam2016}; \citet{CHUBB2018178}; [6] \citet{TASHKUN20111403}; [7] \citet{Sousa-Silva2014}; [8] \citet{Roueff2019}; [9] \citet{VALD};[10] \citet{Kurucz1995}; [11] \citet{ROTHMAN20102139}; [12] \citet{Yurchenko2020}}
    \label{tab:linelists}
\end{table}

\bibliography{main}
\end{document}